\documentclass{aa}  
\usepackage{graphicx}
\usepackage{txfonts}
\begin{document}

   \title{The ASTRODEEP Frontier Fields Catalogues}

   \subtitle{III. Multiwavelength photometry and rest frame properties of  MACS-J0717 and MACS-J1149}

   \author{Di Criscienzo, M.\inst{1}, Merlin E.\inst{1}, Castellano, M.\inst{1}, Santini, P.\inst{1}, Fontana, A.\inst{1}, Amorin, R.\inst{2,3},  Boutsia, K..\inst{4}, Derriere S.\inst{7}, Dunlop, J.S.\inst{5},  Elbaz, D.\inst{6}, Grazian, A..\inst{1}, McLure, R. J.\inst{5}, M\'armol-Queralt\'o, E.\inst{5},  Michalowski, M. J.\inst{5}, Mortlock, S.\inst{5}, Parsa, S.\inst{5}, Pentericci, L.\inst{1}
          }

   \institute{1-INAF -- Osservatorio Astronomico di Roma, Via Frascati 33, 00040, Monte Porzio Catone (RM), Italy
              \email{marcella.dicriscienzo@oa-roma.inaf.it}\\
              2-Cavendish Laboratory, University of Cambridge, 19 JJ Thomson Avenue, Cambridge, CB3 0HE, UK\\
              3-Kavli Institute for Cosmology, University of Cambridge, Madingley Road, Cambridge CB3 0HA, UK\\
              4-Carnegie Observatories, Colina El Pino, Casilla 601 La Serena, Chile
              5-SUPA, Institute for Astronomy, University of Edinburgh, royal Observatory. Edinburgh, EH9 3HJ, UK\\
              6-Laboratoire AIM-Paris-Saclay, CEA/DSM/Irfu-CNRS- Universite\' Paris-Saclay, pt courrier 131, F-91191 Gif-sur-Yvette, France\\
              7-Observatoire astronomique de Strasbourg, Universit\'e de Strasbourg, CNRS, UMR 7550, 11 rue de l'Universit\'e, F-67000,Strasbourg, France\\
             }

   \date{}

 
  \abstract
   {}
   {We present the  multiwavelength photometry of two Frontier Fields massive galaxy clusters MACS-J0717 and MACS-J1149 and their parallel fields, ranging from HST to ground based K and Spitzer IRAC bands, and the public  release  of photometric redshifts and rest frame properties of galaxies found in cluster and parallel pointings. This work was done within ASTRODEEP  project  and aims to provide  a reference for future investigations of the extragalactic populations}
   {To fully exploit the depth of the images and  detect  faint sources we used an accurate  procedure which  carefully removes the foreground light of bright cluster sources and the intra-cluster light thus  enabling detection and measurement of accurate fluxes in crowded cluster regions. This same procedure has been  successfully used to derive the photometric catalogue of MACS-J0416 and Abell-2744.}
   {The obtained multi-band photometry was used to derive photometric redshifts, magnification and  physical properties of sources. In line with  the  first two FF catalogues released by ASTRODEEP, the photometric redshifts  reach $\sim$4$\%$ accuracy. Moreover we extend the presently available samples to galaxies intrinsically as faint as  {\it H}160$\sim$32-34 mag thanks  the magnification factors induced to strong gravitational lensing.  Our analysis allows us to probe galaxy masses   larger then  10$^{7}$ M$\odot$ and/or   SFR=0.1-1M$\odot$/yr  out to redshift z$>6$.}
   {}

   \keywords{catalogs:Methods:data analysis; galaxies:distances and redshifts; galaxies:high-redshift
               }

 \authorrunning{Di Criscienzo et al.} 
   \maketitle
%

\section{Introduction}

 The Hubble Frontier Fields (FF) program \citep{lotz2017}  has been conceived and designed  to explore the highest redshift Universe down to the faintest rest-frame luminosities attainable ahead of JWST, by combining the capabilities of the Hubble Space Telescope (HST) with the amplification power of massive galaxy clusters. The program (PI. Lotz), started in 2012, using HST director discretionary time, has devoted  560 orbits  ($\sim$630 hours) to observe six clusters of galaxies. The FF target clusters were selected as six of the most powerful gravitational lenses presently known, providing lensing amplifications of typically 2 over a significant fraction of the HST/WFC3 field of view up to 10-50 in the most extreme cases.\\
The HST images are supplemented by a wealth of data including  Spitzer and   ground-based  imaging  and  spectroscopic follow-up.\\
The key science driver of the FF programme is shedding light on the properties of galaxies at high redshift (z$>5$),  which are critically important for our understanding of the processes involved in the reionization of the Universe and are presently constrained only from the brightest galaxies discovered in blank-field surveys \citep{castellano2016b,menci2016,bouwens2016,bouwens2016b,mcleod2016,vanzella2017a,vanzella2017b,livermore2017,wei2017}.\\
To achieve the ambitious goal of probing  the distant universe to an unprecedented depth it    is  important  to  develop accurate photometric  procedures that reveal  the power of the deepest images. This is the main scope of the European FP7-Space project ASTRODEEP,  a coordinated and comprehensive program of i) algorithm/software development and testing; ii) data reduction/release, and iii) scientific data validation/analysis of the deepest multiwavelength cosmic surveys\footnote{For more information visit http://astrodeep.eu.}.\\
In the first two papers, \citet{merlin2016} and \citet{castellano2016}, we  described the procedures developed within this collaboration to produce multiband and photometric redshift catalogues and their application to  the first two released  FF Abell-2744 and MACS-J0416.\\
In this paper, we present the public release of the  multiwavelength photometry of MACS-J0717+3745  and MACS-J1149.5+2223 (hereafter M0717 and M1149) that include both Hubble Space Telescope (HST) ACS and WFC3, Keck-MOSFIRE Ks-band and Spitzer-IRAC observations.\\
The paper is structured as follows: in Section 2  we describe the dataset used in this study; Section 3 gives a short description of the procedure we applied  to obtain the detection catalogue  and  photometric measurements in optical and NIR bands. In Section 4   we present the released catalogue   describing in particular   the procedure used to compute the  photometric redshifts, magnification  and rest-frame galaxy properties. Conclusions close the paper.\\
In the following we adopt the $\lambda$-CDM concordance cosmological model(H$_o$ = 70 km/s/Mpc, $\Omega$$_M$ = 0.3 and $\Omega$$_\lambda$ = =0.7). All magnitudes are in AB system unless explicitly mentioned


\section{The dataset}
M0717 and M1149 are the third and the fourth  of a total of six twin  fields observed by HST in seven  optical and  near-infrared bands: F435W, F606W, and F814W from ACS/WFC and F105W, F125W, F140W, and F160W from WFC3/IR. Each of these fields is observed by HST in parallel mode, i.e. cluster and a blank adjacent field.\\
We used the final reduced and calibrated v1.0 mosaics released by STScI, drizzled at 0.06'' pixel-scale. A detailed description of the acquisition strategy and of the data-reduction pipeline can be found in the STScI data release documentation at https://archive.stsci.edu/pub/hlsp/frontier/.
We  also include the  MOSFIRE@Keck  {\it Ks} images from \citet{brammer2016} and the IRAC 3.6 and 4.5  data acquired by Spitzer  under Director Discretionary time (PI Capak).\\
In Table 1 we list PSF FWHM and limiting magnitudes  of the  dataset. For the  HST images the depths have been computed as the magnitudes within a circular aperture of two times the FWHM of 5$\sigma$ detections in the H160 images, as measured by SExtractor on PSF-matched images. To estimate the depths of the  MOSFIRE and IRAC images, we use the corrected RMS maps (see below) computing $f_{5\sigma}= 5 \cdot \sqrt{A_{aper}} \cdot f_{RMS}$ in each pixel, where A$_{aper}$ is the area of a circular region with radius equal to the PSF FWHM,   and taking as final value  the mode of the distributions. 
\begin{table*}
\begin{center}
\caption{PSF FWHM and depths of the dataset (see text).} 
\begin{tabular}{|l||c|c||c|c|}
\hline 
Image &   PSF FWHM('')& Limiting AB magnitude &  PSF FWHM('') & Limiting AB magnitude\\
\hline
\hline 
& \multicolumn{2}{ c|| }{\bf{M0717 Cluster}}& \multicolumn{2}{ c|| }{\bf{M0717 Parallel} }\\
\hline
ACS  {\it B}435  &0.11 &28.64 &0.10&28.71\\
ACS  {\it V}606  &0.13 &28.67 &0.12&28.92\\
ACS  {\it I}814  &0.16 &28.99 &0.14&29.13\\
WFC3  {\it Y}105 &0.16 &29.33 &0.17&28.94\\
WFC3  {\it J}125 &0.18 &28.98 &0.18&28.96\\
WFC3  {\it JH}140&0.18 &29.02 &0.18&28.97\\
WFC3  {\it H}160 &0.18 &29.06 &0.17&28.97\\\
MOSFIRE  {\it Ks}&0.4  &25.08 &0.4&25.19\\
IRAC 3.6  &1.66 &25.47 &1.66&25.22\\
IRAC 4.5  &1.72 &25.22 &1.72&25.19\\
\hline 
& \multicolumn{2}{ c|| }{\bf{M1149 Cluster } }& \multicolumn{2}{ c|| }{\bf{M1149 Parallel}}\\
\hline
ACS  {\it B}435  &0.11 &28.30 &0.10&28.26\\
ACS  {\it V}606  &0.15 & 28.88 &0.10&28.71\\
ACS  {\it I}814  &0.15 &29.08 &0.13&28.90\\
WFC3  {\it Y}105 &0.15 &29.25 &0.17&29.33\\
WFC3  {\it J}125 &0.17 &29.12 &0.16&29.02\\
WFC3  {\it JH}140&0.19 &28.72 &0.17&29.02\\
WFC3  {\it H}160 &0.17 &29.18 &0.17&29.09\\
MOSFIRE  {\it Ks}&0.5  &24.65 &0.5&24.52\\
IRAC 3.6  &1.66 &25.41 &1.66&25.08\\
IRAC 4.5  &1.72 &25.71 &1.72&25.21\\
\hline 
\hline 
\end{tabular}
\end{center}
\label{tabrates}
\end{table*}

\section{Multi-wavelenght photometry}
 
\subsection{Removing the ICL and bright cluster members} 
\begin{figure*}
\centering
\includegraphics[width=14cm]{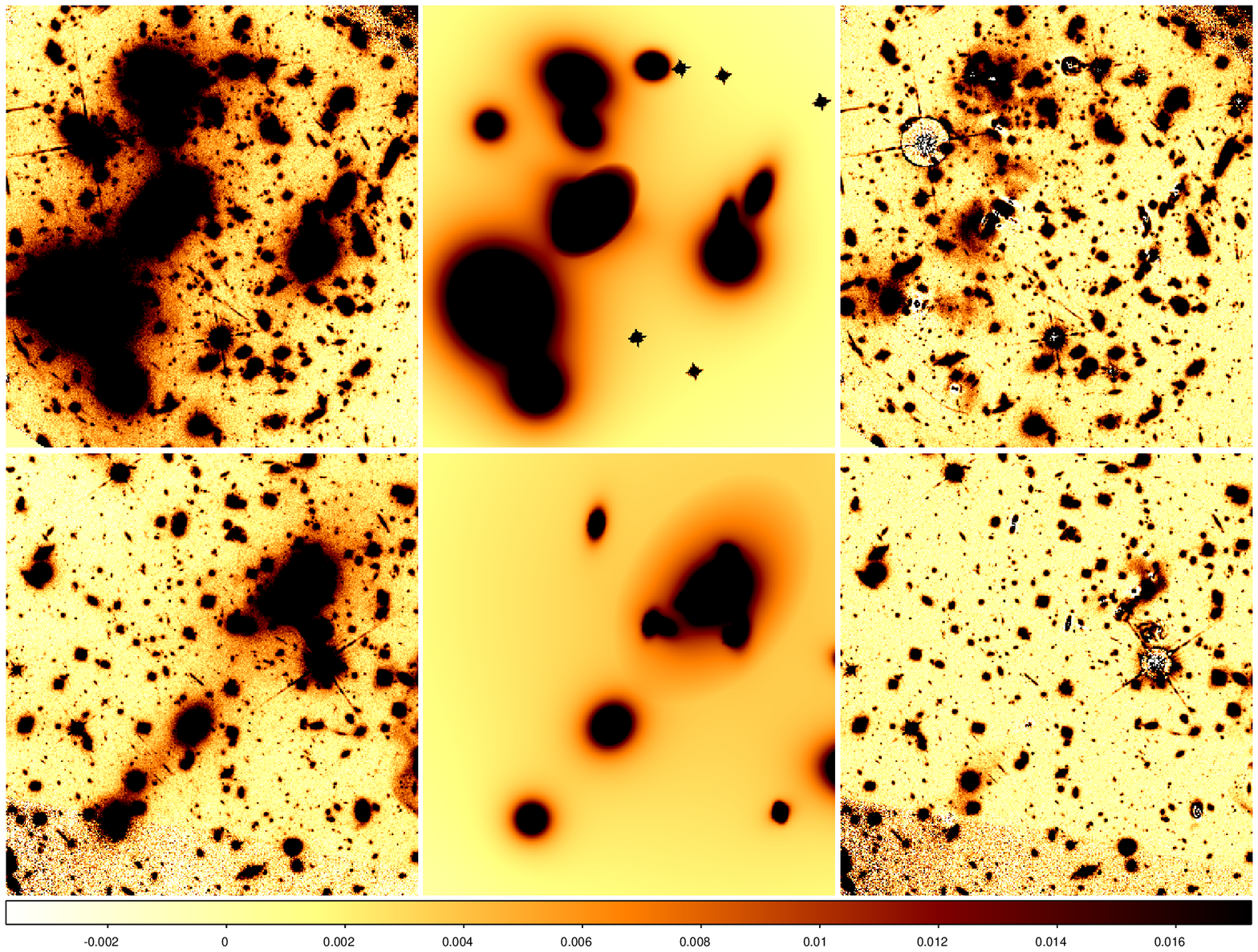}
\caption{
In this figure are illustrated the basic steps performed on the  {\it H}160 images  to remove the light of the cluster sources of M0717(uppper panels) and M1149(lower panels). From the left to the right: original images, Galfit models of bright objects and ICL, and final residual image (=observed - model) after median filtering. All the images are in logarithmic greyscale with the same cuts. 
}
\label{f1}
\end{figure*}

To fully exploit the depth of the images and  detect  faint sources we used an accurate  procedure to remove the foreground light of bright cluster sources and the intra-cluster light (ICL). This procedure is described in details in \citet{merlin2016} and it is even more necessary, compared to  the previously studied FFs, for M0717 and M1149 in which  few multiple merging and subclusters are present making the  ICL bright and patchy\footnote{This step is obvioulsly unecessary in the case of parallel fields.}.
In brief, we initially estimated a first-guess model for the  ICL component  masking  S/N$>$10  pixels  and we fitted the diffuse  light with a Ferrer \citep{binney1987} profile centered on  the mass center of the whole cluster. Then on the ICL-subtracted  {\it H}160 image, with   an iterative method which uses both Galapagos \citep{barden2012} and Galfit\footnote{Galapagos and Galfit are two public  data analysis algorithms that   fit 2-D analytic functions to galaxies and point sources directly to digital images.} \citep{peng2011}, we derived a one/two component  fit   of the brightest cluster galaxies. Finallly these fits are used to refine the model of ICL and  to produce  the  residual image (see Fig.\ref{f1}) where the patchy ICL and the light from  bright sources are subtracted. \\
Unlike in MACS-J0416 and Abell-2744, where all bright galaxies were Galfit-ed with two components to fit the central regions accurately \citep{merlin2016}, here we found a better solution (a flatter residual image) using a single component   (or very faint second one) for those bright galaxies which are located in  the crowed regions (subclusters). Instead we add, during the ICL refinement fit,  a second component to the ICL Ferrer profile centered on the subclusters.\\
In both  clusters  there is a  saturated star in the central  part of the  H image  whose light must be removed to produce  accurate photometry of the faint galaxies. To do this we have subtracted  most of the light from the saturated star using as a PSF the star itself together with three of its rotations of 90,180 and 270 degrees respectively.\\
Finally a median filtering was applied to remove the remaining intermediate scale background residuals. As demonstrated in  \citet{merlin2016} the detection on these  residual image, as opposed to the detection on the original images, allows   a more efficient recovery of the faint sources. \\
We apply the same procedure to all the other HST bands. For consistency and to reduce the computing times, we sequentially move from the {\it H} band to the bluer bands, adopting as first-guess parameter for both ICL and bright cluster galaxies the best-fit parameter of the band immediately redward of it (e.g. we use the {\it H}160  band parameters as first guess to fit the ICL and bright sources in the {\it JH}140 band, those of {\it JH}140  when fitting the  {\it J}125 and so on). 

As a final refining step we used the Galfit model image to estimate the photon noise in each pixel contributed by the bright Galfit-subtracted sourcestaking into account the image exposure time. The resulting ''photon noise image'' was then added 
to the original RMS map summing the variance in quadrature,  to take into account the effect of the subtracted sources on the detection and the flux measurement in the innermost cluster regions.\\
\subsection{Detection catalogue and  HST photometry}  
The  detection catalogue  was produced  in two steps: first using SExtractor \citep{bertin1996} on the processed  {\it H}160 image using a revised  HOT+COLD approch \citep{galametz2013,guo2013} and then adding the additional objects detected in a median average of the  {\it Y}105+ {\it J}125+ {\it JH}140+ {\it H}160 bands which are  undetected  in the H  band
This last step is more effective in the identification of very blue galaxies close to the detection limit of the images, that are expected to include a good fraction of those at redshift 6-8. Table 2 lists the total number of  sources detected  after each step.  In the final  catalogues these IR detected objects are identified  as ID=20000+their original ID.

\begin{table}
\begin{center}
\caption{Total number of  cluster bright objects (N$_{brightobj}$), of detected sources in  {\it H}160 images (N$_{Hdetect}$) and of new sources in  IR stack images (N$_{IRdetect}$).} 
\begin{tabular}{|l|c|l|c|}
\hline
Image &  N$_{bright}$& N$_{Hdetect}$ &  N$_{IRdetect}$\\
\hline 
\hline
M0717cl  &14 &3096&972\\
M0717par  &0 &2181&1266\\
M01149cl  &23 &3379&972\\
M01149par  &0 &2270&1133\\
\hline
\end{tabular}
\end{center}
\label{tabnumber}
\end{table}

The combined detection catalogue was then used to obtain  the photometric measurement (both aperture and total  photometry) in the other HST  bands  using SExtractor  on processed images   convolved to  {\it H}160 resolution (0.18'')  with a convolution kernel obtained taking the ratio of the PSFs of the two images in the Fourier space. 

\begin{figure}
\begin{minipage}{0.48\textwidth}
    \resizebox{1.\hsize}{!}{\includegraphics{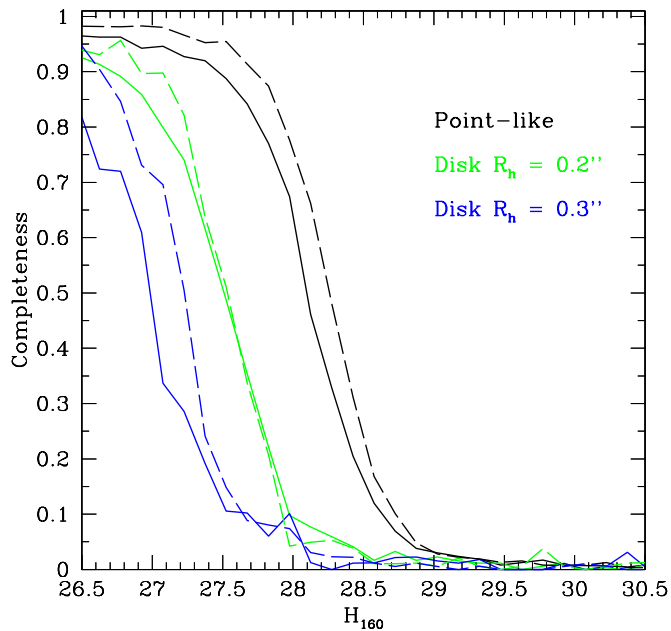}}
\end{minipage}
\caption{Completness of the H-detected catalogue for point-like and disk-like sources in M0717 (solid lines) and M1149 (dashed lines)}
\label{f2}
\end{figure}

We assess the  detection completeness as a function of the H-band magnitude by running simulations with synthetic sources. We first generate populations of point-like  and exponential profile sources, with total H-band magnitude in the range 26.5--30.0 mag. Disc-like sources are assigned an input half-light radius $R_h$ randomly drawn from a uniform distribution between 0.0 and 1.0 arcsec. At each run 200 of these fake galaxies are placed at random positions in our detection image, avoiding positions where real sources are observed on the basis of the original SExtractor segmentation map. We then perform the detection on the simulated image, using the same SExtractor parameters adopted in the real case. Fig.\ref{f2} shows the completeness as a function of the total input magnitude of different  simulated objects (both point- and disk-like). We find that the 90\% detection completeness for the point sources is at H $\sim$ 27.2(27.8) for M0717(M1149) and decreses to H $\sim$ 26.5(26.6) and H$\sim$ 25.7(26.3) for disk-like galaxies of $R_h=0.2$arcsec and $R_h=0.3$ arcsec respectively.

\subsection{K and IRAC Photometry with T-PHOT}
\begin{figure*}
\vskip-10pt
\centering
\includegraphics[width=14cm]{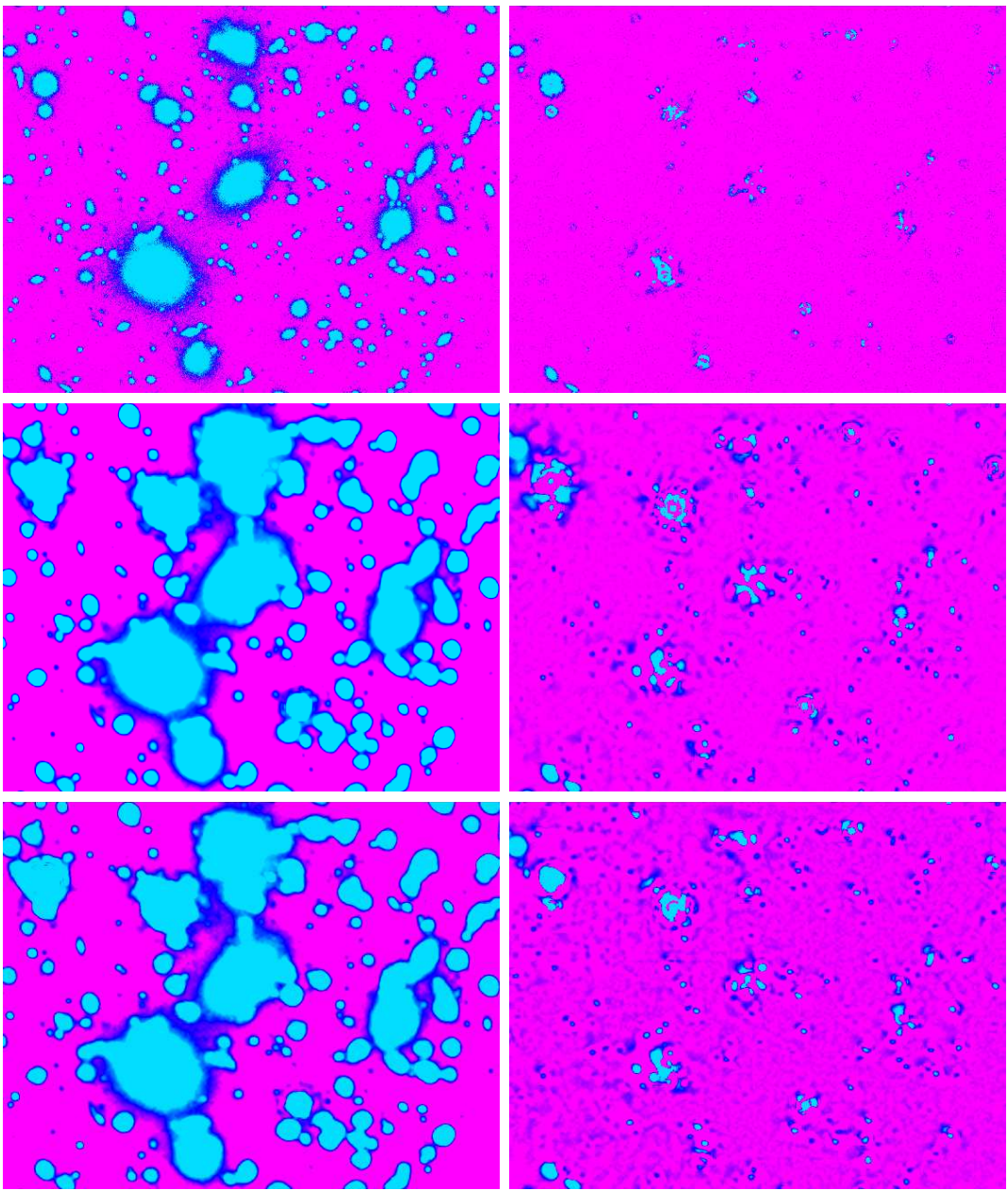}
\caption{Original (left) and residual (right) images in K (upper panels), IRAC-CH1 (central panels) and IRAC-CH2 (lower panels) bands of M0717 after processing with T-PHOT (see Section 3.3 for details).}
\label{f3}
\end{figure*}

K and IRAC photometry  are  obtained via a template-fitting  technique with T-PHOT \citep{merlin2015,merlin2016b} using galaxy shapes in the detection band  {\it H}160 as ``priors''. In this purpose we take  advantage of  T-PHOT V2.0 that allows us to simultaneuosly use as templates the observed galaxy shapes (for all faint objects) and the analytic profiles for the ICL and bright cluster galaxies. In the latter case, after some test, we have decided to fix the ratio between the two components (when present) used for the analytic fits, in order to avoid possible degeneracy issues in the fitting procedure.\\
As discussed at length in \citet{merlin2015}, the segmentation of the objects obtained by SExtractor map may be too small to capture the whole galaxy shape, potentially leading to biases in the flux estimate with T-PHOT. In order to minimize this effect the SExtractor output map has been dilated with the same procedure described in \citet{galametz2013}   before being fed to T-PHOT, enlarging the size of the segmented area of each source by a given factor, depending on the original area.
We have then prepared the measurement image by applying to  the RMS  and background   a corrective factor via injection of fake PSF-shaped sources in about  200 positions  in  empty regions without detected sources.
After having measured the flux of the fake point sources injected at the selected positions, we computed  the RMS map multiplicative factor required to make the distribution of the measured S/N having standard deviation consistent with 1. Instead to derive the correcting factor for the background we measured the shift of the mean of the distribution of the fake sources on copies of the images having small constant artificial background offsets and computing the offset required to make the measured shift consistent with zero. In the case of the  K band images, to take into account the noise correlation  we added a further correcting factor for the background  to be consistent with the magnitudes published in \citet{brammer2016}.\\
Following the procedure used to derive the photometric catalogue of MACS-J0416 and Abell-2744,   we  have also   estimated a local background  for each source and combined  all the measurements to build a global  background image which was then subtracted from the original image. Figure \ref{f3} shows the  residual images obtained  subtracting the scaled models generated by T-PHOT compared with original K and IRAC images of M0717.\\

We follow the same strategy to process the parallel fields, of course without the need to include any analytical model in the priors list.

\section{ Results}
We distribute   final complete multiwavelength photometric catalogues of  four fields (two centered on clusters M07171 and M1149 + two parallel fields)  which contains 10 bands fluxes and magnitudes, and corresponding uncertainties. All the fluxes  were finally  corrected for galactic exinction  derived with Schlegel et al. (1998) dust emission maps. A flag (called RELFLAG) is associated to each object, that gives indication of the robustness of photometric estimates. ``Good sources'' have RELFLAG=1 which means they have more than 5 HST bands with reliable (Sextractor's internal {\tt FLAGS$\le$16}) flux measurement available.
As in \citet{castellano2016} we complement the publicly released catalogues with photometric redshift, stellar mass and star formation rate as described below.
\subsection{Photometric redshifts and comparison with spectroscopic samples}
\begin{figure*}
\centering
\includegraphics[width=8cm]{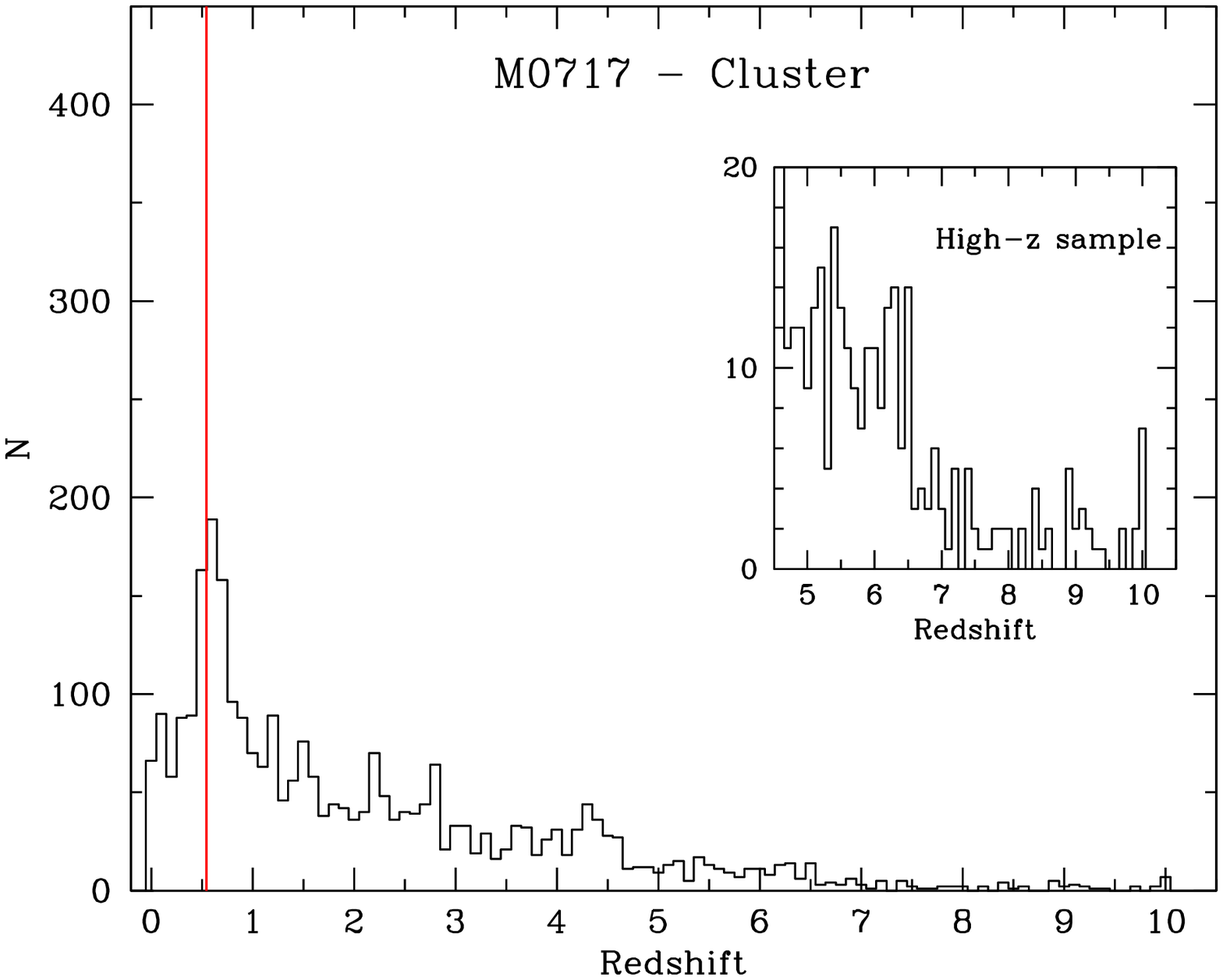}
\includegraphics[width=8cm]{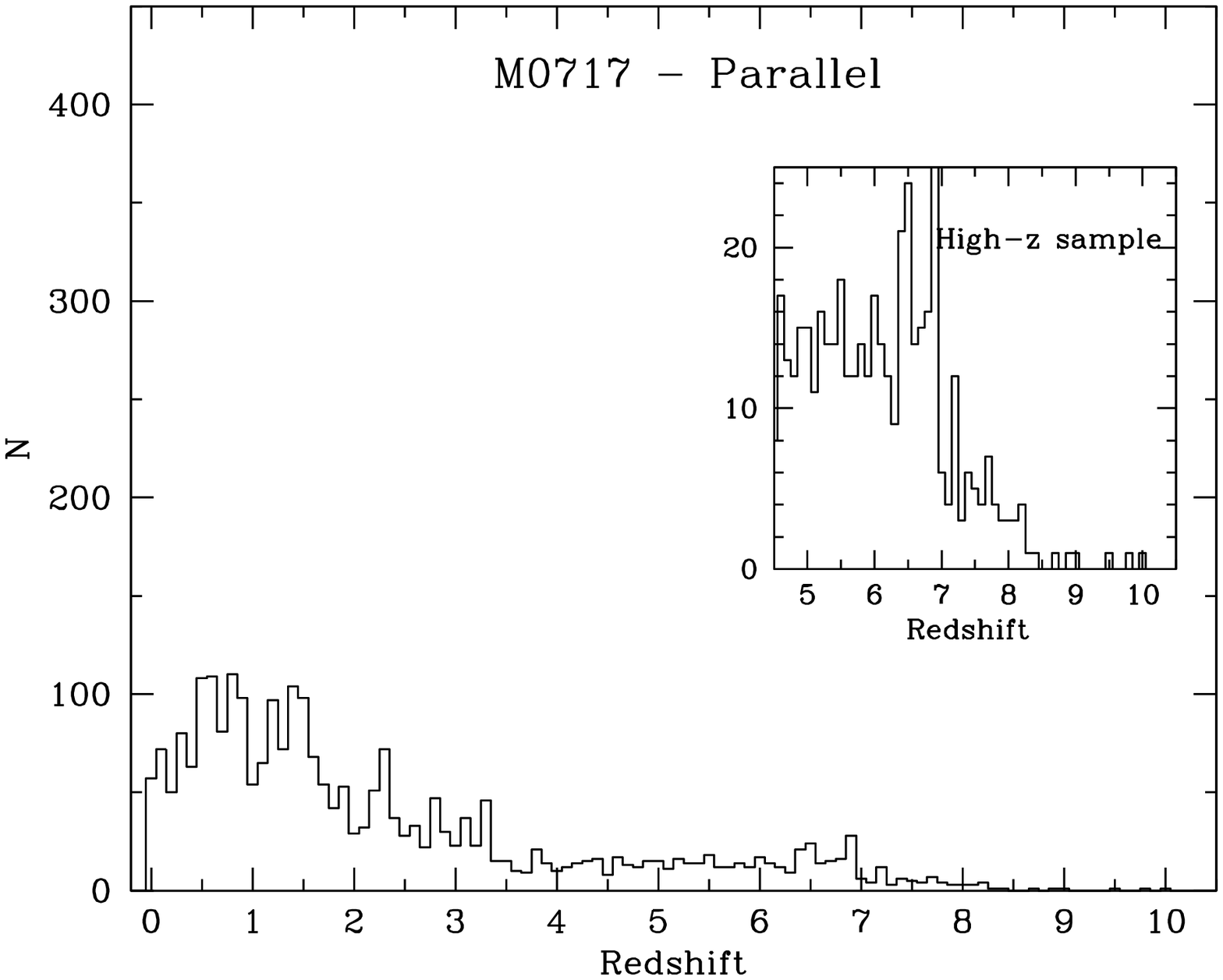}
\includegraphics[width=8cm]{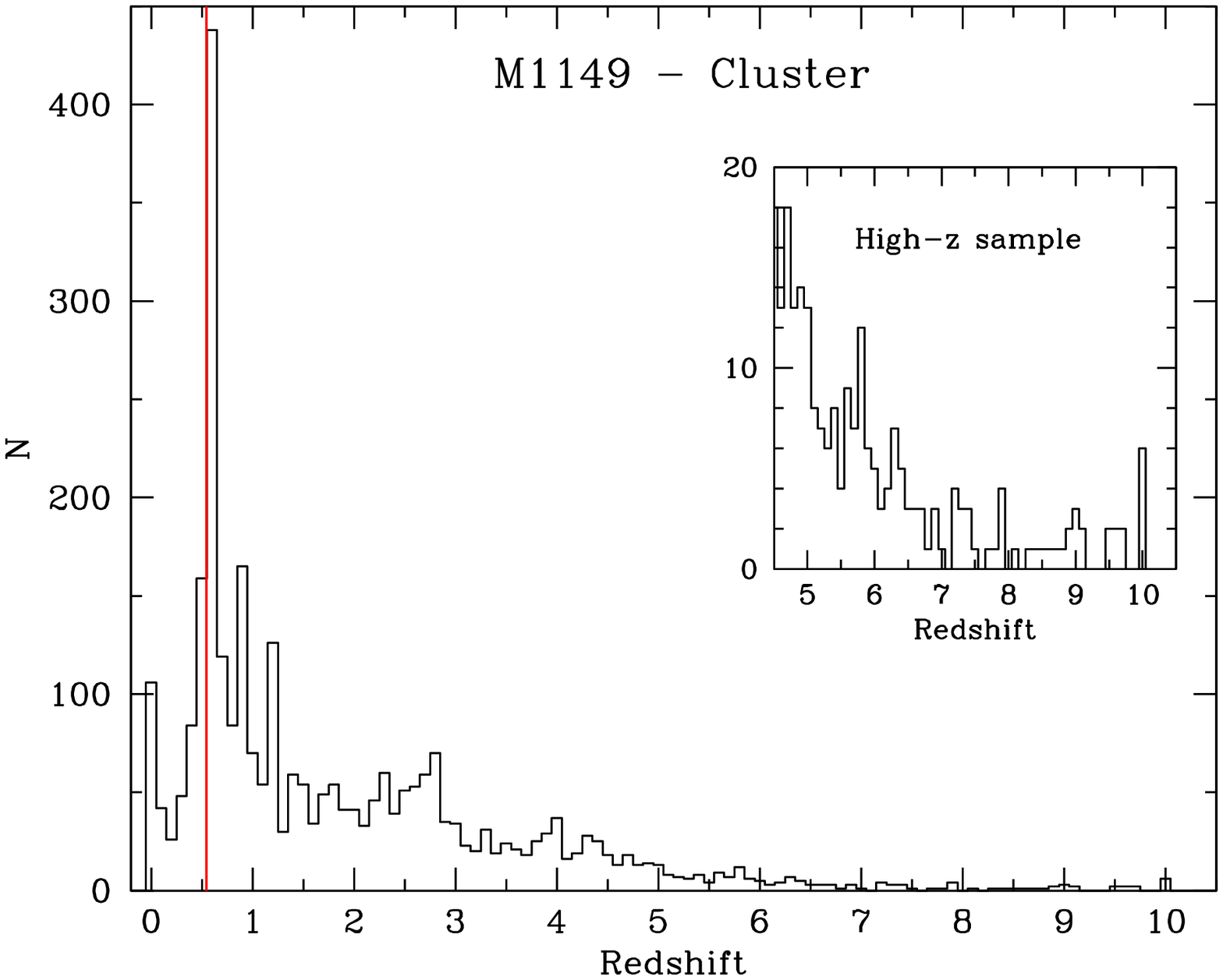}
\includegraphics[width=8cm]{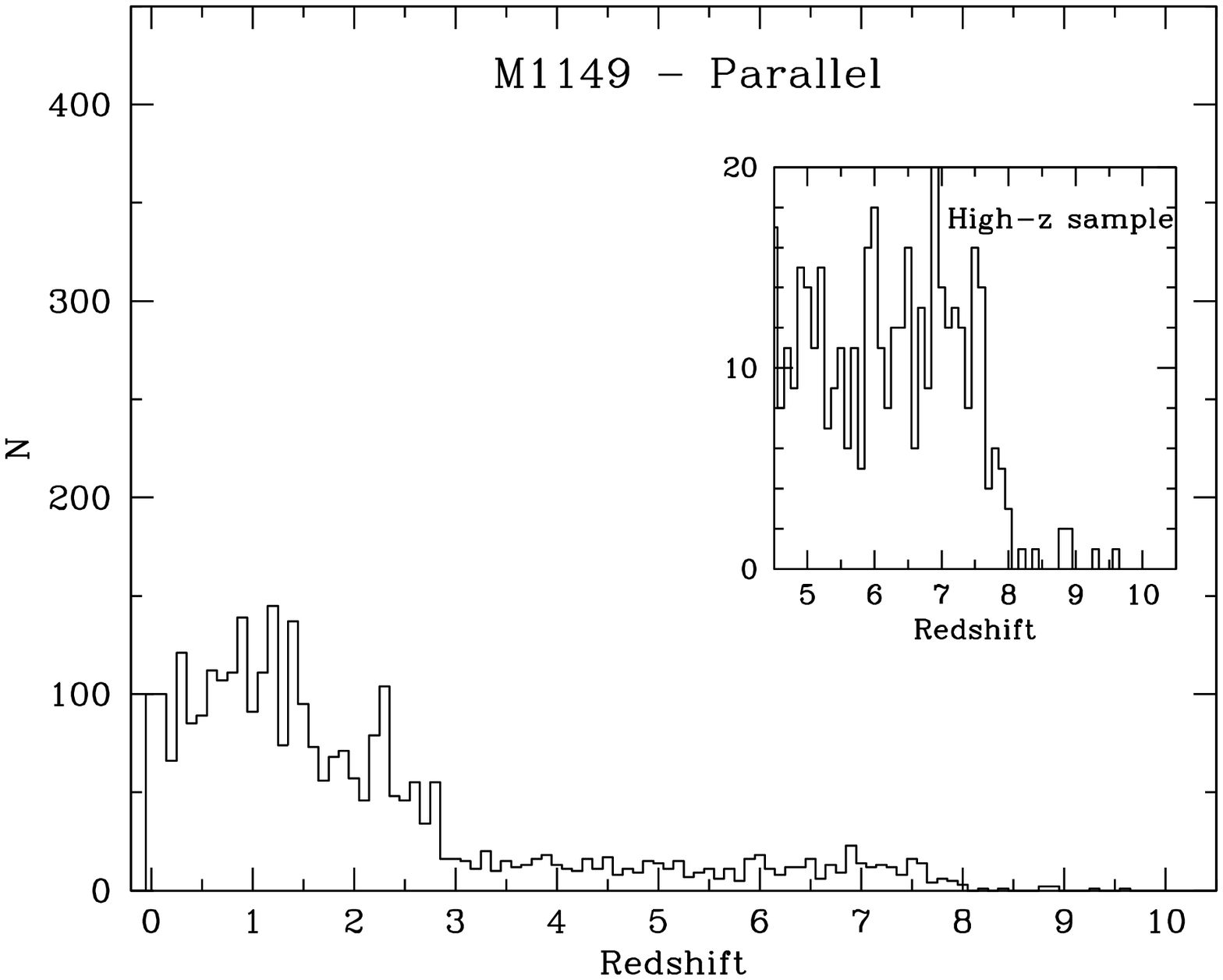}
\caption{Photometric redshift distribution in our four catalogues. Insets show a zoom for object with $z>5$  in order  to appreciate the high redshift tail of the distribution.}
\label{zphot}
\end{figure*}

To minimize systematic  effects due to the use of a single method we have measured photometric redshifts  using six different algorithms:  1) {\tt OAR} \citep{castellano2016}; 2) {\tt McLure}\citep{mclure2011}; 3){\tt Mortlock }\citep{arnouts1999,blanton2007}; 4){\tt Parsa }\citep{arnouts1999,ilbert2006}; 5){\tt Marmol-Queralto-1 }\citep{brammer2008,blanton2007}; 6){\tt Marmol-Queralto-2}\citep{brammer2008,fioc1997}). 
These techniques  are described in detail in Section 3 of \citet{castellano2016}. Photometric redshifts are determined for all ''good sources'' using all available bands with the exception  of K and  IRAC fluxes which are unreliable due to severe blending with other sources (T-PHOT parameter MaxCvRatio $>$1.0, see Merlin et al. 2016). In  Fig. \ref{zphot} we  show  the resulting  median  photometric redshifts  distribution computed for all ''good sources''.

 Objects having a positive match (within 1 arcsec) with reliable public spectroscopic  samples are assigned the measured spectroscopic redshift. In particular, we consider the redshifts from the public dataset  by \citet{ebeling2014}, GLASS (Treu et al. 2015, for sources with quality flag Q=3 and Q=4) together with the arcs from \citet{limousin2012} in the case of M0717 and \citet{smith2009} for M1149. 

\begin{figure*}
\begin{minipage}{0.48\textwidth}
    \resizebox{1.\hsize}{!}{\includegraphics{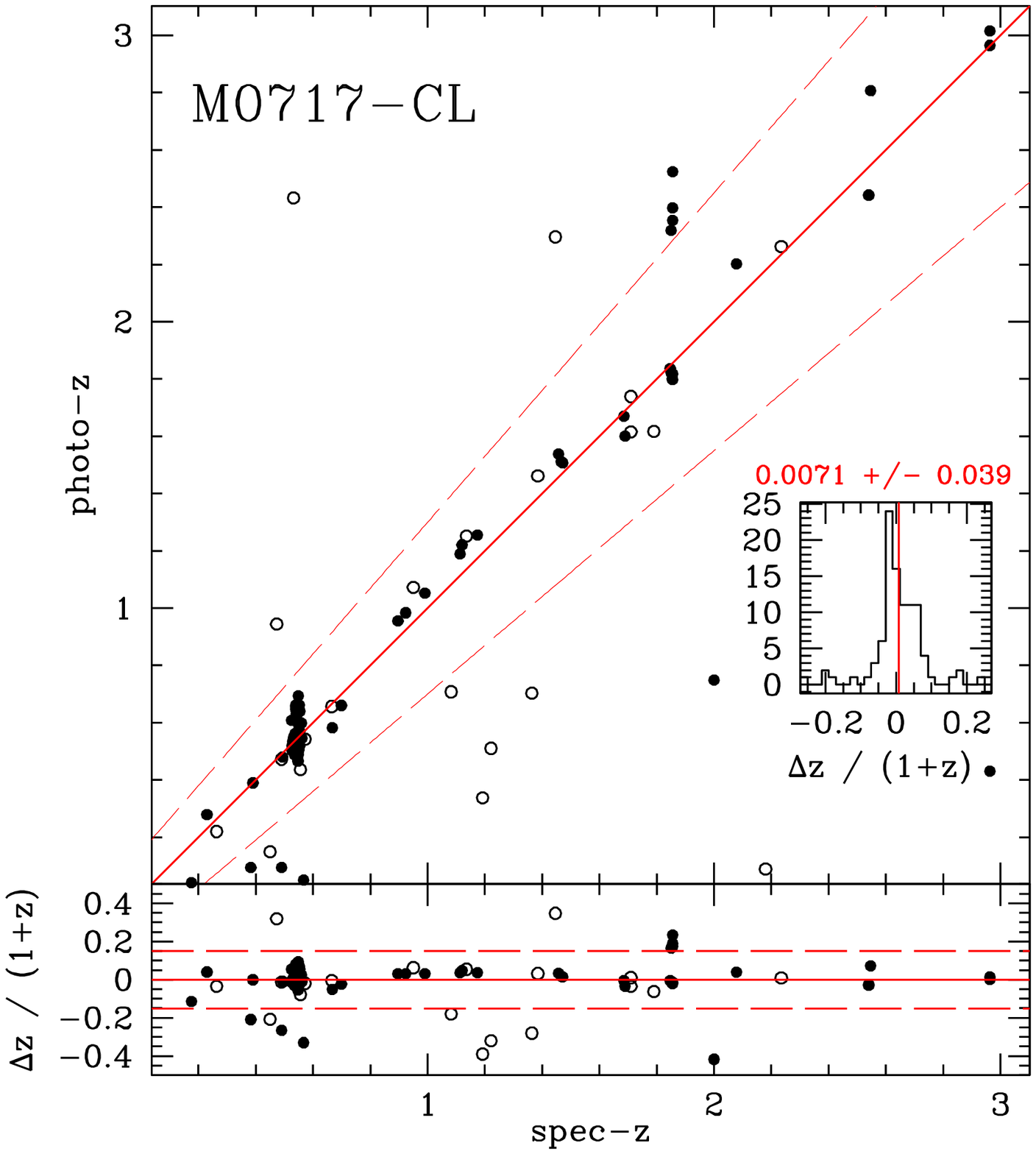}}
\end{minipage}
\begin{minipage}{0.48\textwidth}
\resizebox{1.\hsize}{!}{\includegraphics{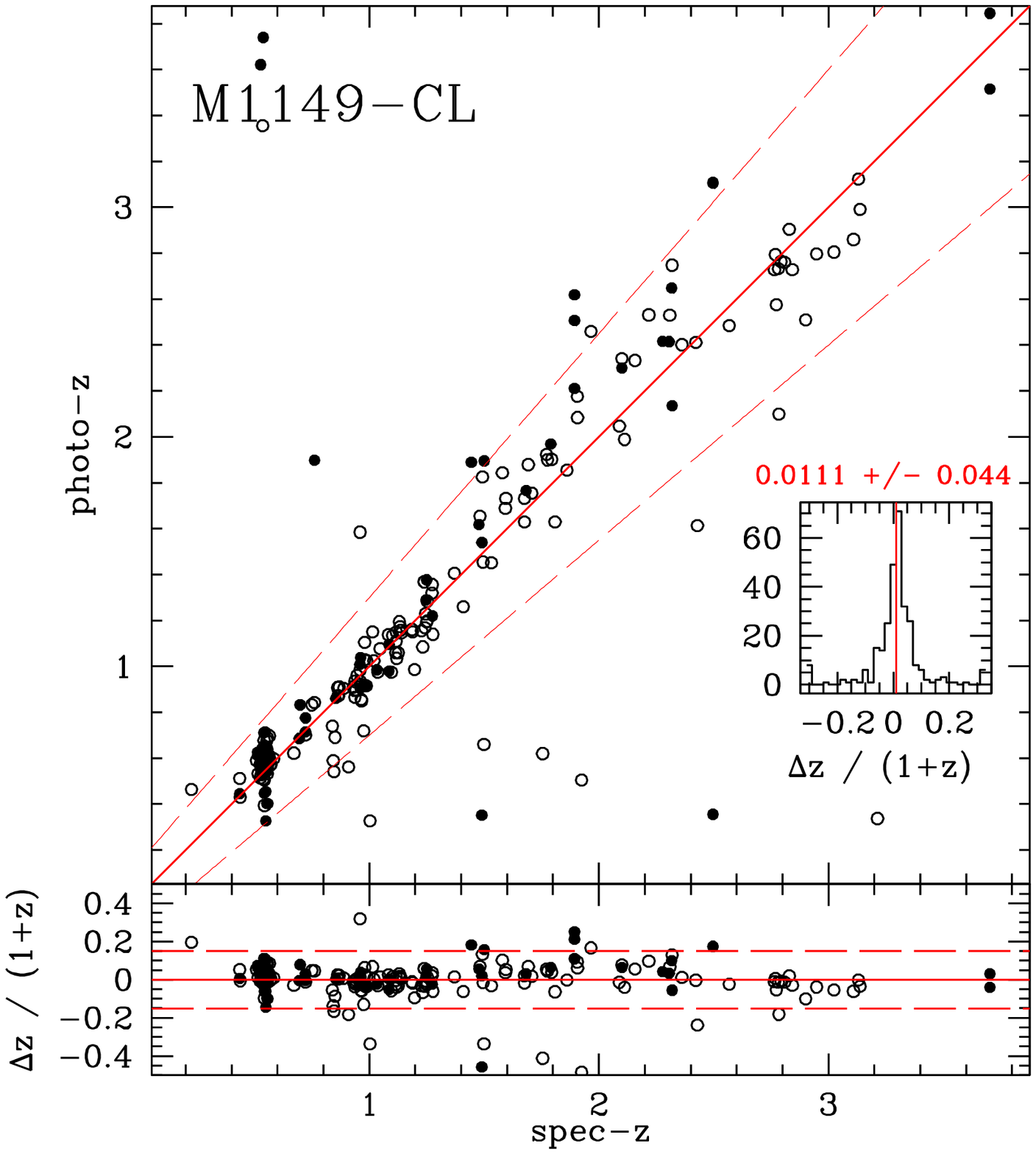}}
\end{minipage}
\caption{Comparison between photometric median redshifts of our  good sources (RELFLAG=1) and the spectroscopic estimate for M0717 (left) and M1149 (right). Filled circles  represent best quality spectroscopic redshifts (Q=4) . In the lower panels  we show  $\Delta$z / (1+z)=(z$_{spec}$-z$_{phot}$)/(1+z$_{spec}$) as a function of the spectrosocpic redshift. In the inner small panels  the distribution of $\Delta$z / (1+z) is shown together with   its average (vertical line) and rms after excluding ''outliers'', as discussed in the text}
\label{fzspec}
\end{figure*}

When compared with spectroscopic results,  median values of photometric redshift  are more accurate than the individual runs  computed with the six different techniques ( 0.046 $\le$ r.m.s $\le$ 0.055) and for this reason we give the median value in the released catalogue. In Fig.\ref{fzspec} we show the comparison  between our median estimate of photometric redshift  and spectroscopic value  for   all our ''good source''  in the cluster's field. Following Dahlen et al 2013 we define as outliers all sources having |$\Delta$z| / (1+z)=|(z$_{spec}$-z$_{phot}$)|/(1+z$_{spec}$)$\ge$0.15. 
In Table 3 we report the number of outliers and the statistic  in each cluster.
In the case of the parallel fields  the final sample includes only two  objects with spectroscopic redshift and it makes no sense to provide the statistics.


\begin{figure}
\begin{minipage}{0.48\textwidth}
\vskip-100pt
\resizebox{1.\hsize}{!}{\includegraphics{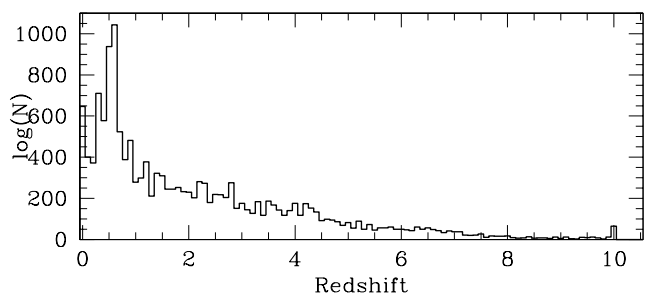}}
\end{minipage}
\caption{Distribution of photometric redshift of all good sources detected in the four clusters}
\label{allz}
\end{figure}

\begin{table}
\begin{center}
\caption{Photomeric redshift accuracy.}
\begin{tabular}{|c|c|c|c|c|}
\hline
Field &  Spec. sample & outliers & $<$$\Delta$z/(1+z)$>$ & $\sigma_{\Delta z/(1+z)}$\\
\hline 
\hline
M0717  &109 &18\% &0.0071 & 0.037\\
M1149  &285 & 9\% &0.011 &0.044\\
\hline
\end{tabular}
\end{center}
\label{tabnumber}
\end{table}
In Fig. \ref{allz} we show the distribution of photometric redshift for all objects of the first four FFs (Castellano et al. 2016 + this work).

\subsection{Demagnified  number counts and rest -frame physical properties}
\begin{figure*}
\begin{minipage}{0.48\textwidth}
\resizebox{1.\hsize}{!}{\includegraphics{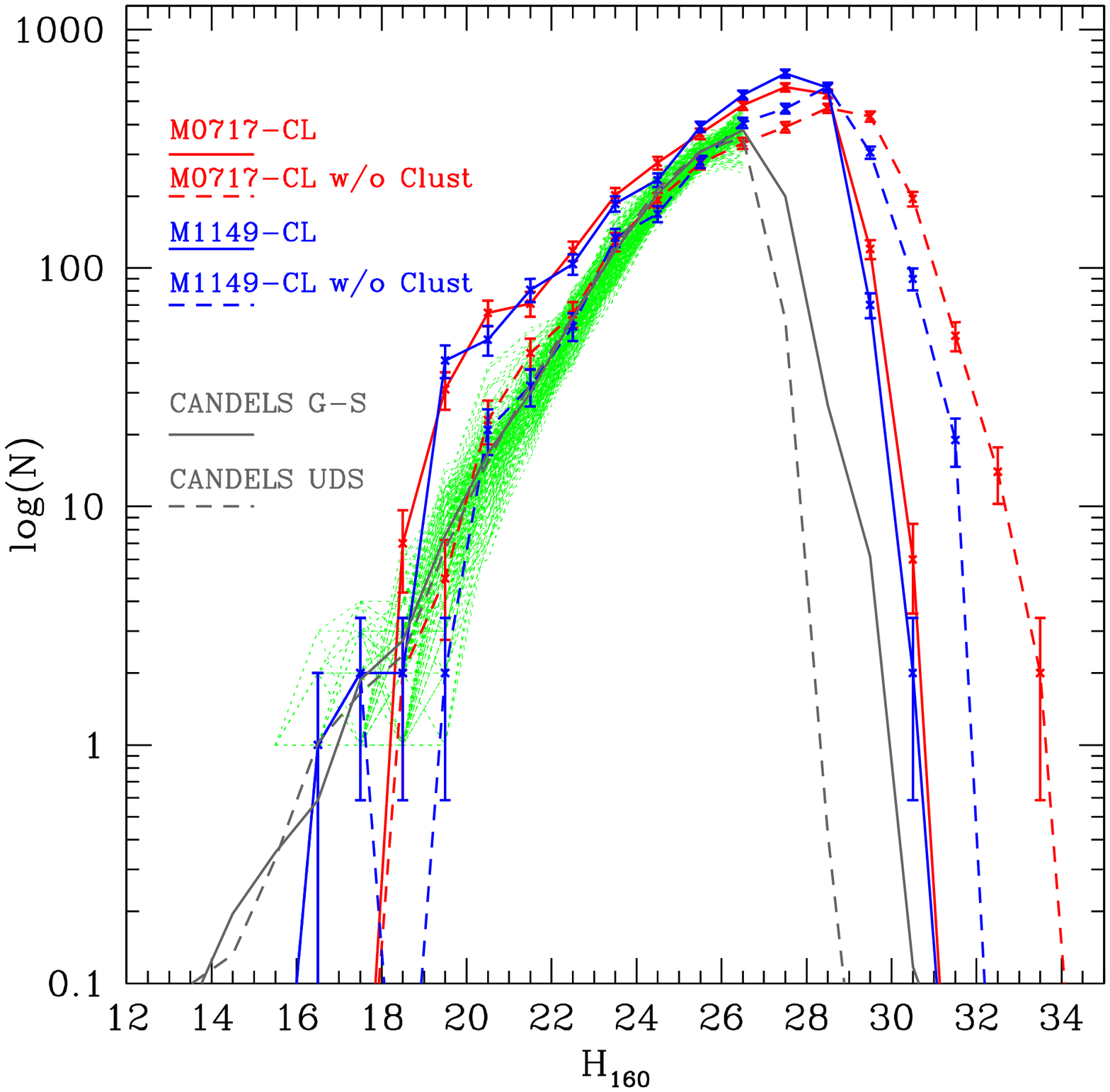}}
\end{minipage}
\begin{minipage}{0.48\textwidth}
\resizebox{1.\hsize}{!}{\includegraphics{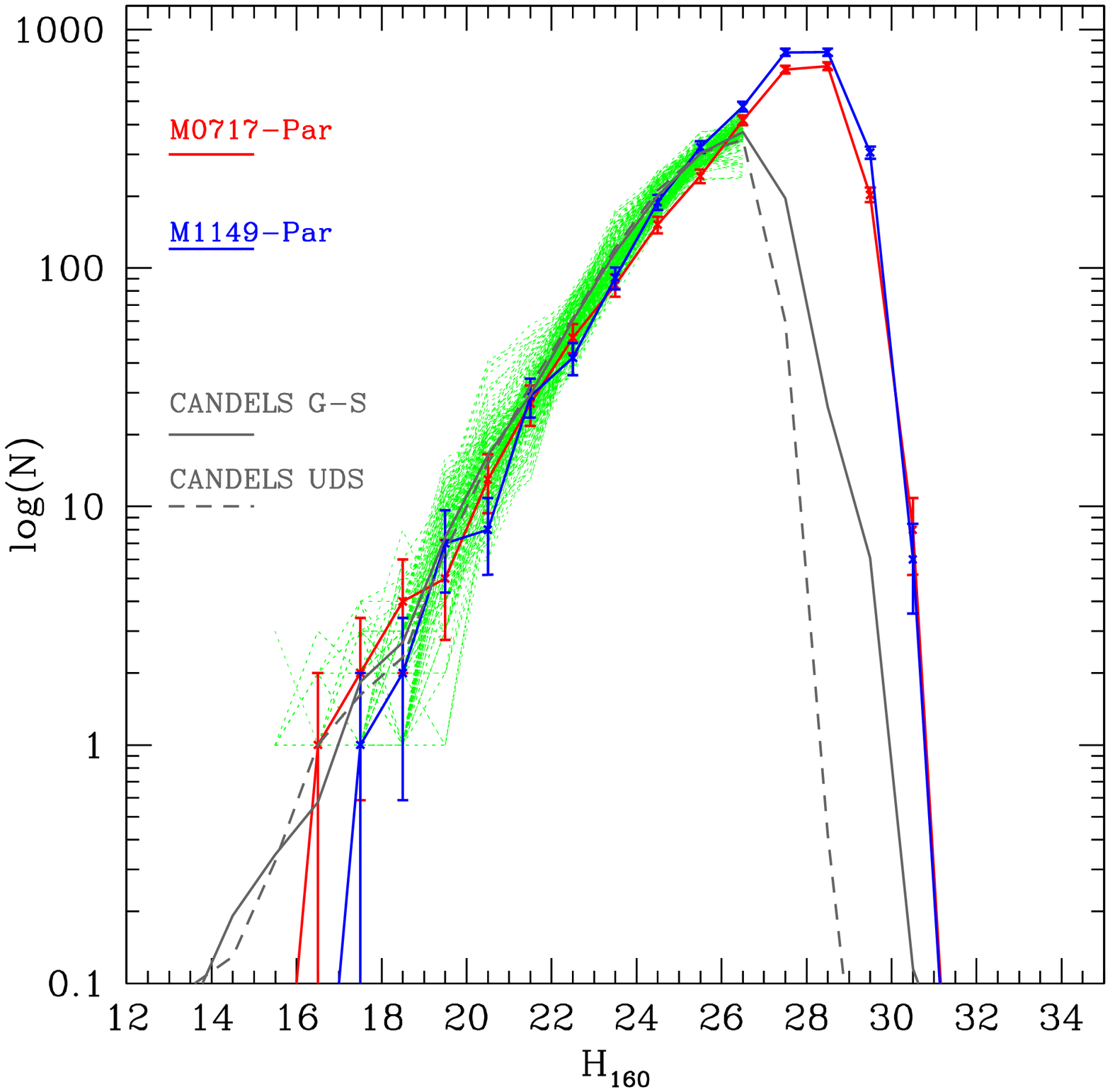}}
\end{minipage}
\caption{(Solid lines)Demagnified number counts in the cluster fields when sources with z$_{phot}$  within 0.1 the redshift of the  relative cluster are removed. As a comparison, number counts normalized to the same  FF area from the public CANDELS GOODS-South and UDS catalogues  are shown. The green lines in particular are number counts from randomly chosen portions   having the same area of the FF pointings. }
\label{f7}
\end{figure*}

\begin{figure}
\begin{minipage}{0.48\textwidth}
\resizebox{1.\hsize}{!}{\includegraphics{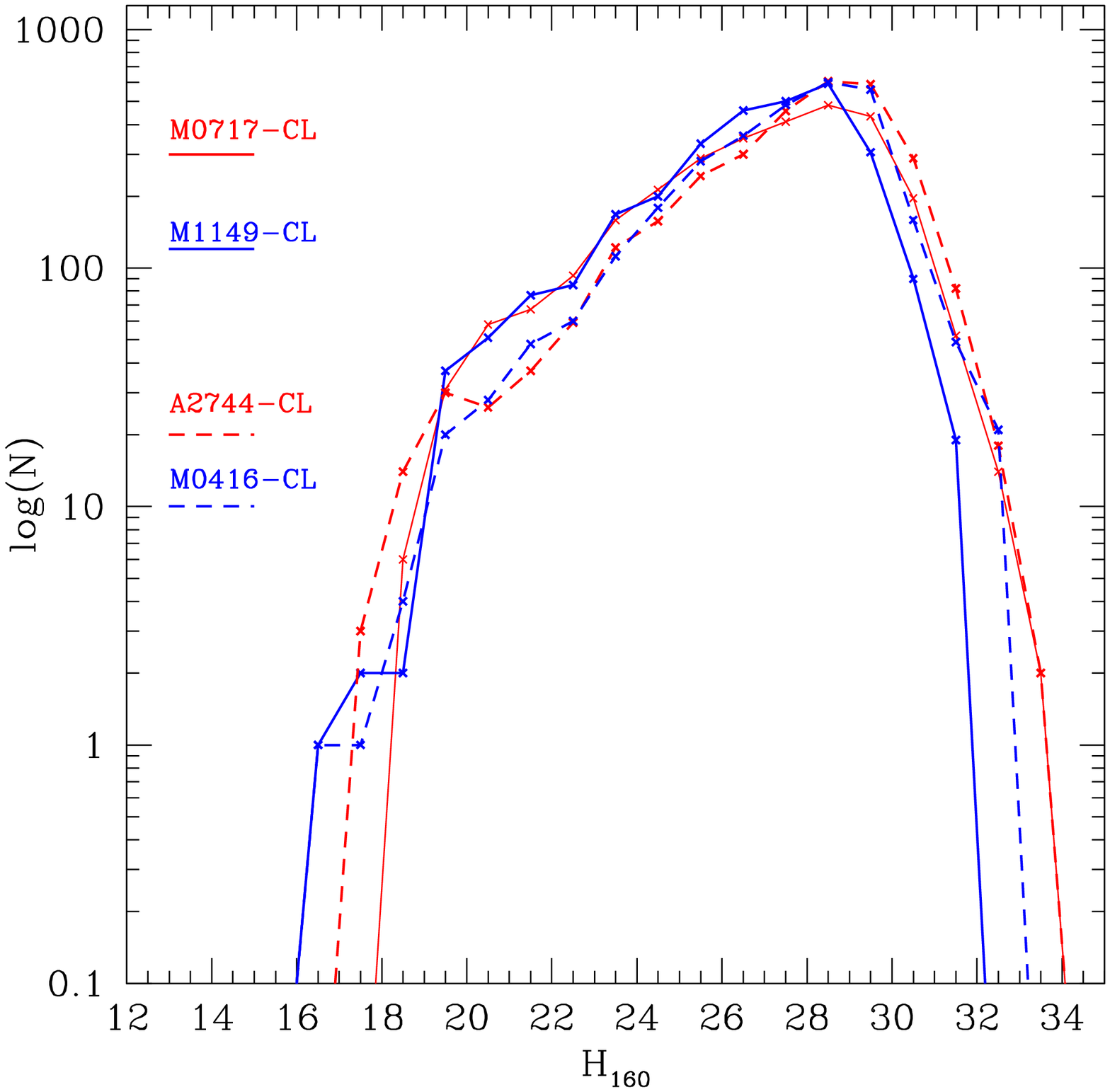}}
\end{minipage}
\caption{Demagnified number counts in the cluster fields investigated in this work compared with the previous two  FFs from  Castellano et al.(2016).}
\label{f7bis}
\end{figure}

Ultra deep IR observations of the FF  in combination with the strong gravitational lensing effect allow us to probe stellar masses and star formation rates at unprecedented low limits. 
We have first determined magnification values from all available lensing models  described in detail  on the FF website\footnote{http://www.stsci.edu/hst/campaigns/frontier-fields/Lensing-Models}, on an object-by-object basis taking into account source position and redshifts.We assign a magnification to each source in our catalogues as the median values computed  using  the available lensing models.
The magnified number  counts are shown in Fig.\ref{f7}  compared with  total number counts from CANDELS GOODS-South  \citet{guo2013} and UDS  \citep{galametz2013} surveys normalized to FF area. For magnitudes brighter then  {\it H}160=26mag,  the  number counts are consistent  with the CANDELS ones once magnification is taken into accounts and when sources with z$_{phot}$  within 0.1 the redshift of the  relative cluster are removed. At fainter magnitudes  the FF cluster pointings allow us to detect sources up to 4 magnitudes intrinsically fainter than objects in the deepest areas of the CANDELS fields. Figure \ref{f7bis} shows the comparison with Abel-2744 and MACS-J0416.

Finally we also release de-magnified M$_{star}$ and SFRs as a function of redshift for galaxies in our catalogues obtained through SED-fitting.
Galaxy properties are  computed by fitting Bruzual \& Charlot (2003) templates with our custom {\tt zphot.exe} code \citep{giallongo1998,fontana2000,grazian2006} at the previously determined median photometric redshift. In the BC03 fit we assume exponentially decling star formation histories with e-folding time 0.1 $\le$ $\tau$ $\le$ 15, a Salpeter (1955) initial mass function and we allow both \citet{calzetti2000} and Small magellanic Cloud \citep{prevot1984} extinction laws. We fit all the sources both with stellar emission templates only and including the contribution from nebular continuum and line emission following Schaerer $\&$ de Barros (2009) under the assumption of an escape fraction of ionizing photons fesc= 0.0 (see also Castellano et al. 2014). 
The Frontier Fields allow us to probe the galaxy distribution down to very low masses and SFRs, including objects with M$_{\star} \sim$\,10$^{7}$ M$_{\odot}$ and SFR$\sim$\,0.1-1 M$_{\odot}$\,yr$^{-1}$ at $z>6$, depending on magnification. 

\section{Conclusions}
We have presented the public release of  multi-wavelenght photometry of the Frontier Fields M0717 and M1149 (cluster and parallel pointings) including optical and NIR ACS and WFC3, MOSFIRE Ks and IRAC 3.6 and 4.5 IRAC bands. We have followed the same method used and described in detail in Merlin et al. 2016 for Abell-2744 and MACS-J0416 with small differences mainly due to the extreme  crowding of the two investigated clusters. The catalogues also report first high-level data products such  as  photometric redshifts, magnification factors and rest frame properties for the detected objects, which  can be downloaded from the ASTRODEEP website at http://www.astrodeep.eu/ff34 or at http://astrodeep.u-strasbg.fr//ff//index.html. 
This work, as it is happening the first two papers (see for example Vanzella et al. 2017a,b)  aims to provide a reference for future investigations of the extragalactic populations.
\begin{acknowledgements}
      The authors acknowledge the contribution of the FP7 SPACE project ASTRODEEP (Ref.No:212725) supported by European Commission. R.A. acknowledges the support from the ERC Advanced Grant 'QUENCH'.
\end{acknowledgements}

\end{document}